\documentstyle[twoside,fleqn,espcrc2,epsfig]{article}

\newcommand{\AmS}{{\protect\the\textfont2
  A\kern-.1667em\lower.5ex\hbox{M}\kern-.125emS}}

\title{The Massive Schwinger Model in a Fast Moving Frame}

\author{H. Jirari\address{D\'epartement de Physique,
        Universit\'e  Laval, Qu\'ebec, 
        Qu\'ebec G1K 7P4, Canada},
        H. Kr\"oger$^{\rm a}$,
        K.J.M. Moriarty\address{
        Department of Mathematics, Statistics 
        and Computing Science,     
        Dalhousie University,
        Halifax, Nova Scotia B3H 3J5, 
        Canada}       
        and N. Scheu$^{\rm a}$}

\begin{document}

\newcommand{\pstxt}[3]{\begin{figure}[b]
\psfig{figure=#2,width=\linewidth,angle=-90,bbllx=19pt,bblly=9pt,bburx=592pt,bbury=783pt}
\caption[#1]{{\small\bfseries #1}. {\footnotesize #3}}\label{#2}
\end{figure}
}

\newcommand{\psnew}[1]
{\psfig{#1,angle=-90,bbllx=19pt,bblly=9pt,bburx=592pt,bbury=783pt}}



\begin{abstract}
We present a non-perturbative study of the massive Schwinger model.
We use a Hamiltonian \\ approach, based on a momentum lattice 
corresponding to a fast moving reference frame, and equal time 
quantization.
\end{abstract}


\maketitle

\section{INTRODUCTION}
The most important non-perturbative approch to study $QCD$ is lattice gauge theory. However, there are some observables, where computational progress
in lattice gauge theory has been slow. Examples are: Finite density thermodynamics (quark-gluon plasma), exited states of the mass spectrum (hadrons and mesons), dynamical scattering calculations of cross sections and
hadron structure functions, in particular in the region of small $Q^2$ and small
$x_B ~ (10^{-2}$  to $10^{-5})$. A non-perturbative Hamiltonian approach may be a viable alternative. The Hamiltonian method does have some advantages over the 
Lagrangian method. It is relatively simple to obtain wavefunctions of the 
hadronic states, and it is likely that some physical information on the
glueballs, e.g., the wavefunction can be derived in this formulation more easily \cite{xiang96,xiang97}.
In deep inelastic lepton-hadron scattering, the success of the
parton model suggests the physical idea to use a fast moving frame also for a
computational study of those processes. The parton model can be justified using the operator product expansion. In equal-time quantization, the Breit-frame is the most convenient choice of frame in order to interpret structure functions.
In Ref.\cite{Kroger97} the authors have proposed such a scheme based on
equal-time quantization, using a lattice Hamiltonian on a momentum lattice
corresponding to a fast moving frame (Breit-frame). It has been applied to the scalar $\phi^{4}$ theory in $3+1$ dimensions. Distribution functions and the mass spectrum in the close neighbourhood of the critical line of the second order phase transition have been computed.
Here we study a simple gauge theory, namely $QED_{1+1}$ \cite{Kroger98}, the so-called massive Schwinger model,
in this framework \cite{schwinger62}. The purpose of this work is to show:
(a) The use of a fast moving frame, such that $v < c$, in conjunction with {\em
equal-time} quantization works well also in
$QED_{1+1}$. We obtain quite precise results in the
ultra-relativistic region $m/g \to 0$.
(b) We consider as most important new results of this work the non-perturbative
results of the dependence of the vector and scalar mass on the $\theta$-angle.

\section{Method}
Starting from the Lagrangian, we use the axial gauge $A^{3}=0$ to obtain the
Hamiltonian
\begin{eqnarray}
\lefteqn{H = \int_{-L}^{L} dx^{3}
(\bar{\psi} \gamma^{3} i \partial_{3} \psi + \bar{\psi} \psi )}\nonumber\\
& & + \frac{g^{2}}{2} \int_{-L}^{L} dx^{3} 
(\psi^{\dagger}\psi) \frac{1}{-\partial^{2}_{3} } (\psi^{\dagger}\psi).
\end{eqnarray} 
One introduces a space-time lattice given by spacing $a$ with
$N=\frac{2L}{a}$ lattice nodes. Via discrete Fourier transformation one goes
over to a momentum lattice, with cut-off $\Lambda = { \pi \over a}$ and
resolution
$\Delta p = { \pi \over L}$. Motivated by the parton picture we make the
assumption that left-moving particles are not dynamically important, if
physical particles are considered from a reference frame
\begin{figure}[htb]
\psnew{file=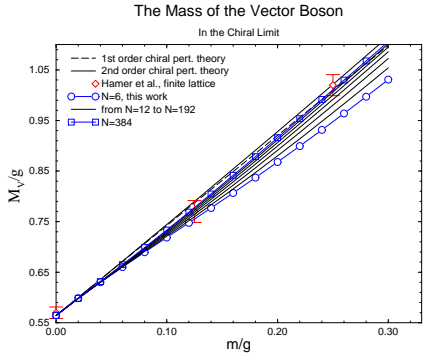,width=0.7\linewidth}
\caption{The dimensionless mass of the vector boson versus $m/g$ in the chiral region.}
\label{fig:massvector}
\end{figure}
\begin{figure}[htb]
\psnew{file=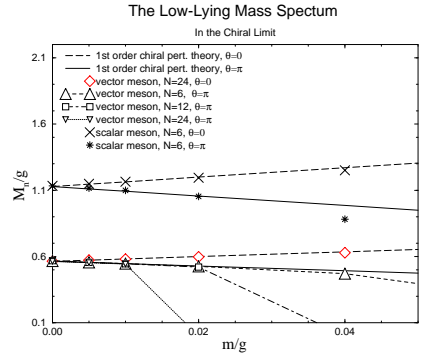,width=.7\linewidth}
\caption{Dependence of the dimensionless mass of vector and scalar particle
on the $\theta$-angle after subtraction of $\theta$-vacuum energy.}
\label{fig:thetadependence}
\end{figure}
characterized by a velocity $v= { P \over E }$
which is close to the velocity of light. We thus consider a momentum lattice
where $0 \leq p^{0}, p^{3} \leq P$.
In order to minimize the number of virtual particle pairs created
from the vacuum, we choose a small lattice size. The reason for this
is that the number of vacuum pairs is roughly proportional
to a vacuum density times the lattice size.
On the other hand, a fast-moving physical particle is Lorentz contracted;
thus it fits into a small lattice volume (when compared to the rest-frame).
For the purpose of computing the mass spectrum, we need to determine
the vacuum energy. Because the vacuum has the quantum number $\vec P=0$,
the vacuum energy (and only this) is computed in the rest frame.
Because the model is super-renormalizable, one can perform the continuum limit
$a \to 0$. The only renormalization necessary is the subtraction of the vacuum
energy. On a space-time lattice, one has to satisfy a physical condition
(scaling window)  $a < \xi a < L$, where $\xi$ is the correlation length in
dimensionless units, related to the physical mass of the ground state by $M = {
1 \over \xi a}$.
In a strongly relativistic system, $M \ll P$,
thus when $a \to 0$ the scaling window is replaced by
$1/P < 1/M < L$.
For more details compare with Ref.\cite{Kroger97}.

\section{Numerical results}
\subsection{Mass spectrum}
We diagonalize the Hamiltonian in a sector with momentum $\vec{P} = 0$ to
obtain the vacuum energy $E_{vac}$.
Then we diagonalize the Hamiltonian in a sector $\vec{P} \neq 0$ corresponding
to a relativistic velocity. This yields an energy spectrum $E'_{n}$.
The physical energies are obtained from $E_{n} = E'_{n} - E_{vac}$. The mass
spectrum is then given by $M_{n} = \sqrt{ E^{2}_{n} - \vec{P}^{2} }$.
The low lying states of the massive Schwinger model are a vector state and next
a scalar state \cite{Hamer97}.
The mass of the vector boson in the chiral region is shown in Fig. \ref{fig:massvector}.
The vector particle is almost entirely a fermion-antifermion ($q\bar{q}$) bound state.
We find good agreement with chiral perturbation theory \cite{Adam96} and
with finite lattice results by Hamer et al.\cite{Hamer97}.

\subsection{Dependence on $\theta$-angle}
The massive Schwinger model has $\theta$-vacua and one can study
its $\theta$-action. The wavefunction is invariant under local gauge transformations. The $\theta$-angle characterizes the behavior of the 
wavefunction under global gauge transformations $\Psi[A] \to e^{i n \theta} \Psi[A]$, where $n=0, \pm 1, \pm 2, \cdots $. 
The mass spectrum of low-lying states as a function of the $\theta$-angle
is shown in Fig. \ref{fig:thetadependence}. The results are in agreement 
with first oder chiral perturbation theory in the ultra-relativistic regime ($m/g < 0.04$ for $N=6$ and $m/g < 0.01$ for $N=24$). The Feynman-Hellmann theorem relates the fermion condensate $<\bar{\psi} \psi >$ to the
derivative of the vacuum energy.
Similarly holds for the vector state
\begin{equation}
< \bar{\psi} \psi > =
\left. \frac{M^{(0)}}{2 \pi} \frac{\partial }{\partial m } M_{v} \right|_{m=0}
\end{equation}
Extracting the slope $\partial M_{v} /\partial m$ from our data
we obtain $< \bar{\psi} \psi > /g = 0.16 \cos(\theta)$ while the exact solution of the massless Schwinger model gives a factor $\frac{e^{\gamma}}{2\pi \sqrt{\pi}} \approx 0.1599$ on the r.h.s.

\subsection{Distribution functions}
In the Hamiltonian approach it is easy to compute the wavefunction of a
low-lying state.
>From the wavefunction one can obtain information on its parton structure,
i.e., the number of partons and their momentum distribution.
The distribution function of the vector boson is given by
\begin{equation}
f(x_{B}) =
 <\Psi_{v}(P) \mid
   {1\over 2}
   \left[ b^{\dagger}_{k} b_{k}
        + d^{\dagger}_{k} d_{k}
  \right]\mid \Psi_{v}(P) >,
\end{equation}
where $x_{B} = k/P$ is the fraction of momentum of the vector boson carried by
the parton, i.e., 
\begin{figure}[t]
\psnew{file=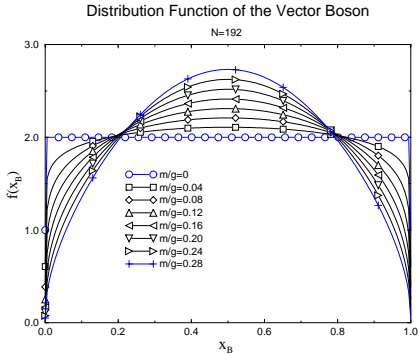,width=0.7\linewidth}
\caption{Distribution function of the vector boson for different $m/g$ in the chiral region.}
\label{fig:distribfct}
\end{figure}
In Fig. \ref{fig:distribfct} we display lattice results of the distribution function
for $m/g = 0$ to $0.28$. In the massless limit the distribution function has the shape of a box. 
Our results are compared with variational calculations using the infinite-momentum 
frame by Bergknoff \cite{Bergknoff77} and also with those by Mo and Perry \cite{Mo93} 
using the light-cone. The most sensitive region is the ultra-relativistic region.
We find agreement in shape with Mo and Perry's results and very good agreement 
with Bergknoff's results.

\section{Summary}
\noindent In this work we have applied a Hamiltonian lattice approach in a fast moving reference frame to study the massive Schwinger model.
We find that the method works well for the computation of the low-lying mass
spectrum and distribution functions. It works also in the presence of the $\theta$-action. Here we have investigated only $\theta=0,\pi$. 
More $\theta$-angles can be studied, e.g. by adding a small number of negative momentum states to the basis.


\begin{thebibliography}{9}
\bibitem{xiang96}X.Q. Luo et al., Mod. Phys. Lett. A11 (1996) 2435;
\bibitem{xiang97}X.Q. Luo et al., Nucl. Phys. B(Proc. Suppl.)53 (1997) 243.
\bibitem{Kroger97}H. Kr\"oger and N.Scheu, Phys. Rev. D56 (1997) 1455.
\bibitem{Kroger98}H. Kr\"oger and N.Scheu, Phys. Lett. B429 (1998) 58.
\bibitem{schwinger62}J. Schwinger, Phys. Rev. 128 (1962) 2425.
\bibitem{Hamer97}C.J. Hamer, Z. Weihong and J. Oitmaa, Phys. Rev. D56 (1997) 55.
\bibitem{Adam96}C. Adam,  Phys. Lett. B382 (1996) 383.
\bibitem{Bergknoff77}H. Bergknoff, Nucl. Phys. B122 (1977) 215.
\bibitem{Mo93}Y. Mo and R.J. Perry, J. of Comput. Phys. 108 (1993) 159.





\end{thebibliography}
\end{document}